\documentclass[prb,nobalancelastpage,nofootinbib,superscriptaddress,showpacs,noeprint,twocolumn]{revtex4-1}

\usepackage{physics}
\usepackage{graphicx}
\usepackage{bm}
\usepackage{bbm}
\usepackage[bb=boondox]{mathalfa}
\usepackage{nicefrac}
\usepackage{subcaption}
\usepackage{verbatim}
\usepackage{scrextend}
\usepackage{tikz}
\usepackage{bbold}
\usepackage{lipsum}

\usepackage[justification=justified,
   format=plain]{caption}

\usepackage{amsfonts}
\usepackage{amssymb}
\usepackage{amsmath}
\usepackage{epsfig}
\usepackage{enumerate}
\usepackage{multirow}

\usepackage{caption}
\usetikzlibrary{positioning}
\usetikzlibrary{fit}
\usetikzlibrary{calc,shadings}
\usetikzlibrary{patterns}

\graphicspath{{figs/}}

\begin{document}

\title{Approximating observables on eigenstates of large many-body localized systems}

\author{Abishek K. Kulshreshtha}
\affiliation{%
	Rudolf Peierls Centre for Theoretical Physics, Clarendon Laboratory, Parks Road, Oxford OX1 3PU, United Kingdom.
}
\author{Arijeet Pal}%
\affiliation{%
	Rudolf Peierls Centre for Theoretical Physics, Clarendon Laboratory, Parks Road, Oxford OX1 3PU, United Kingdom.
}
\affiliation{Department of Physics and Astronomy, University College London, Gower Street, London WC1E 6BT, United Kingdom}
\author{Thorsten B. Wahl}
\author{Steven H. Simon}
\affiliation{%
 Rudolf Peierls Centre for Theoretical Physics, Clarendon Laboratory, Parks Road, Oxford OX1 3PU, United Kingdom.
}%

\date{\today}

\begin{abstract}
Eigenstates of fully many-body localized (FMBL) systems can be organized into spin algebras based on quasilocal operators called l-bits. These spin algebras define quasilocal l-bit measurement ($\tau^z_i$) and l-bit flip ($\tau^x_i$) operators. For a disordered Heisenberg spin chain in the MBL regime we approximate l-bit flip operators by finding them exactly on small windows of systems and extending them onto the whole system by exploiting their quasilocal nature. We subsequently use these operators to represent approximate eigenstates. We then describe a method to calculate products of local observables on these eigenstates for systems of size $L$ in $O(L^2)$ time. This algorithm is used to compute the error of the approximate eigenstates.           
\end{abstract}

\maketitle

\section{Introduction}

Thermalization behavior of closed quantum systems has received heightened interest since the suggestion that Anderson localization could be generalized to systems of interacting particles, a phenomenon dubbed \textit{many-body localization (MBL)} \cite{Nandkishore2015,Alet}. Over the last decade and a half, an increasingly large body of proof has spoken to the existence and complexity of this behavior. Perturbative arguments \cite{Basko2006a,Gornyi2005}, studies using exact diagonalization \cite{Pal2010,Oganesyan2007}, and further mathematical proofs \cite{Imbrie2016a} have all emerged over this period. This body of work has firmly proved the existence of MBL at strong disorder in one-dimension without broken time-reversal symmetry or spin-orbit coupling, while localization at weaker disorder closer to the phase transition is still the subject of exploration \cite{Vosk2015,Potter2015,Khemani2016b}.  Experiments using cold atoms and trapped ions have also revealed robust MBL behavior \cite{Schreiber2015,Smith2016}. 

General many-body states are expressed in a Hilbert space that grows exponentially with system size; MBL systems have the added property of a description using an extensive set of \textit{$l$-bits}\cite{Huse2014,Serbyn2013,Ros2015}, which can be thought of as quasi-local generalizations of physical spins. The $l$-bit algebra implies a set of quasi-local operators: $\tau^z_i$ which measures the $l$-bit on the $i^{\text{th}}$ site and $\tau^x_i$ which flips the $l$-bit on the $i^{\text{th}}$ site. For the spin-$1/2$ systems with which we work, the $l$-bit algebra can be thought of as akin to the Pauli spin algebra. $\tau^z_i$ returns a phase ($\pm 1$) when applied to the eigenstate.

Several algorithms exist to construct these integrals of motion approximately \cite{Pekker2016,Rademaker2017,Chandran2015,Chandran2015b,Pollmann2016,Pekker2017a,Wahl,Thomson2018}, which produce operators that do not commute exactly with the Hamiltonian. Recent algorithms have also been proposed to construct integrals of motion exactly \cite{Kulshreshtha,Goihl2017}.

Our focus in this paper will be on using $l$-bit algebras to construct approximate eigenstates on large MBL systems. Several methods for constructing eigenstates on large MBL systems already exist. One particular class that has shown great success is algorithms based on the density matrix renormalization group (DMRG) algorithm \cite{Schollwock2011}. While DMRG itself finds the ground state of a generic local Hamiltonian, algorithms like shift-and-invert matrix product states (SIMPS) \cite{Yu2015}, DMRG-X \cite{Khemani2016} and En-DMRG \cite{Lim2016}, reviewed further in Section \ref{subsec:current}, are able to compute excited states of MBL systems by exploiting the area-law nature of MBL eigenstates.

Additionally, a class of recent tensor network algorithms \cite{Wahl,Pollmann2016,Pekker2017a,Chandran2015b} provide effient methods of constructing matrices whose columns are approximate eigenstates of the system. These algorithms, also reviewed further in \ref{subsec:current}, use layers of local unitaries to generate a large unitary matrix that approximately diagonalizes the Hamltonian.

We introduce a novel type of algorithm to construct approximate eigenstates on large MBL systems, using exact $l$-bit algebras on small subsystems to approximate an $l$-bit algebra on a larger system. Using this algorithm, eigenstates can be targeted by their $l$-bit labels, allowing one to access any eigenstate in practice. Further, we extend the algorithm to show how it can also be used to measure expectation values of products of local observables on eigenstates of large systems.

We begin by describing the disordered Heisenberg spin chain and $l$-bit algebras in further detail. We then review the existing classes of methods used to construct approximate eigenstates on large MBL systems. 

In Section \ref{sec:methods}, we describe a new algorithm based on ideas similar to the authors' work in Ref. [\onlinecite{Kulshreshtha}] but far more efficient. The algorithm is used to construct exact $l$-bit algebras on small systems and is labelled \textit{operator localization optimization (OLO)}. We subsequently describe how we use the improved algorithm repeatedly to construct approximate eigenstates on large MBL systems, which we label the $\tau_x$ \textit{network representation} due to its similarity to the tensor network. We additionally introduce an algorithm, labeled the \textit{inchworm algorithm}, to measure products of local observables on the approximate eigenstates. 

In Section \ref{sec:results}, we first test the quality of our eigenstates by measuring their energy fluctuations and compare these results to the tensor network method of Ref. [\onlinecite{Wahl}]. We additionally show how the algorithm can be used to find correlations over large distances. Finally, we conclude the paper by discussing our algorithm in relation to tensor network-class and DMRG-class algorithms and considering future directions.

\section{Phenomenology and other methods}

\label{sec:phenom}

\subsection{XXZ Spin Chain}

We make use of the disordered XXZ spin chain, with the Hamiltonian
\begin{equation}
\label{eq:main_ham} 
H = \sum_{i = 1}^{L-1} \mathbf{S}_i \cdot \mathbf{S}_{i+1} + \sum_{i = 1}^L h_i S^z_i, 
\end{equation}
where $\mathbf{S}_i = \frac{1}{2} \boldsymbol{\sigma}_i$ and the values $h_i$ are drawn randomly and independently from a uniform distribution $[-W,W]$. The properties of this model are well studied. For small $W$, the model is known to obey the Eigenstate Thermalization Hypothesis, while for large $W$, the model is known to exhibit localized, non-ergodic behavior. This behavior can be probed in a variety of ways, including through the level statistics of the energy spectrum \cite{Pal2010}; through the entanglement characteristics of eigenstates and mobility edge \cite{Luitz2015,Bauer2013}; through the behavior of integrals of motion \cite{Kulshreshtha}; and through diffusion characteristics \cite{Agarwal2015}.
Though system behavior for large $W$ and small $W$ are well understood, the crossover between thermal and localized characteristics is still being probed. 

Starting from $W \ll 1$, behavior is thermal. As $W$ increases, Griffiths regions, rare insulating areas surrounded by regions with metallic behavior, begin to dominate the behavior of the system and transport of conserved quantities becomes subdiffusive \cite{Agarwal2015,Vosk2015,Potter2015}. Finally, as $W$ exceeds $W_c$, the system becomes completely localized. 

In the localized phase, eigenstates no longer obey ETH, level statistics display Poisson behavior \cite{Pal2010}, and system entanglement grows as the logarithm of time following a quantum quench \cite{Bardarson2012,Serbyn2013a}. Numerical simulations give estimates for the transition disorder strength at $W_c \approx 3.5$ in the thermodynamic limit\cite{Agarwal2015,Pal2010,Kulshreshtha,Luitz2015}, though this value is subject to finite-size effects. The work of this paper occurs well within the localized regime.

\subsection{Current eigenstate approximation methods}
\label{subsec:current}

Several methods exist to approximate eigenstates on large MBL systems. We provide a brief review of two of these methods here in order to provide context for the method we present in this paper.

The tensor network method for approximating eigenstates, developed and presented in Refs. [\onlinecite{Pollmann2016}] and [\onlinecite{Wahl}], uses layers of small unitary matrices to represent a large unitary matrix that approximately diagonalizes the Hamiltonian. Ref. [\onlinecite{Pollmann2016}] makes use of two-site unitary matrices stacked in multiple layers. Starting from arbitrary unitary blocks, the algorithm sweeps across the system, using the conjugate gradient method to minimize the total variance of the energy of the approximate eigenstates generated by the matrices. The computational cost of this algorithm scales linearly as a function of system size and exponentially as a function of number of layers. Closer to the MBL transition, more layers are required to accurately represent the system's eigenstates as their entanglement properties become less local.

In Ref. [\onlinecite{Wahl}], the scaling of the computational cost is reduced by instead fixing the number of layers at $2$ and increasing the size of the smaller unitary blocks. Using unitary matrices that act on a larger number of sites, fewer layers are required to represent eigenstates to the same accuracy. Ref. [\onlinecite{Wahl}] also makes use of a cost function whose computational cost scales less quickly than that of finding the total eigenstate energy variance. We benchmark our algorithm against this one in Section \ref{sec:results}.

The other method we highlight is DMRG-X, presented in Ref. [\onlinecite{Khemani2016}]. The original DMRG method makes use of the fact that ground states of one-dimensional systems can be represented accurately through matrix product states (MPS) \cite{Verstraete2006,Perez-Garcia2007}. The DMRG method starts from a random matrix product state, sweeping through the system and updating the constituent matrices of the MPS by minimizing an effective Hamiltonian with respect to individual parts of the MPS. 

The DMRG-X method of Ref. [\onlinecite{Khemani2016}] makes use of the fact that eigenstates of MBL systems can be represented efficiently through MPS \cite{Friesdorf2015}. Eigenstates can be targeted by their overlap with physical spin product states. \mbox{DMRG-X} starts from an initial physical spin product state and updates the constituent matrices of the MPS by replacing them with maximally overlapping eigenstates of an effective Hamiltonian, allowing one to target eigenstates based on proximity to a physical spin structure. Another variation of DMRG is presented in Ref. [\onlinecite{Lim2016}]. In this method, labelled En-DMRG, one can target eigenstates by energy using the DMRG and Lanczos methods. As opposed to DMRG, DMRG-X and En-DMRG allow one to target eigenstates across the energy spectrum as long as they can be accurately represented using MPS. 

In Section \ref{sec:concl} of this paper, we describe how our method complements the methods described above.

\section{Methods}

\label{sec:methods}

\subsection{Introduction}
As previously described, eigenstates of FMBL systems can be expressed through an $l$-bit spin algebra, akin to the physical spin algebra. In the absence of spin-spin interaction, eigenstates are physical spin product states and the $l$-bit algebra is simply that of the physical spins. In the presence of spin-spin interactions, the $l$-bit label of an eigenstate corresponds to a quasi-local measurement on the system. In an FMBL system, the weight of the $l$-bit measurement on a site decays exponentially with distance from the site, where weight is defined below. Where the physical spin measurement operator on site $i$ is labelled $\sigma^z_i$, the $l$-bit measurement operator on site $i$ is labelled $\tau^z_i$. The non-trivial action of the $\tau^z_i$ operator on site $j$ decays as $e^{|i-j|/\xi}$, where $\xi$ is the localization length.

Just as the $\sigma^x_i$ operator flips the spin of a state on site $i$, a quasi-local $l$-bit flip operator $\tau^x_i$ can be defined on FMBL systems. The $l$-bit flip operator on site $i$ takes one eigenstate to another whose label $\tau^z_i$ is flipped on site $i$ and is unchanged everywhere else.

$\tau^x_i$ operators can be constructed as 
\begin{equation}
\tau^x_i = U \sigma^x_i U^\dagger ,
\end{equation} 
where $U$ diagonalizes the Hamiltonian and contains the eigenstates of the Hamiltonian in its columns. However, there exist exp($L$) real choices for $U$ for a system of size $L$. These choices correspond to rearranging the columns of $U$ and assigning each column a phase of plus or minus one for a real Hamiltonian. Only certain choices of $U$ will yield quasi-local operators with exponentially decaying weight. Therefore, given the set of eigenstates of a Hamiltonian, finding the correct $U$ is a non-trivial combinatorial optimization problem.

\subsection{Constructing quasilocal operators on small subsystems}
\label{subsec:olo}

\begin{figure}
	\includegraphics[width=\linewidth]{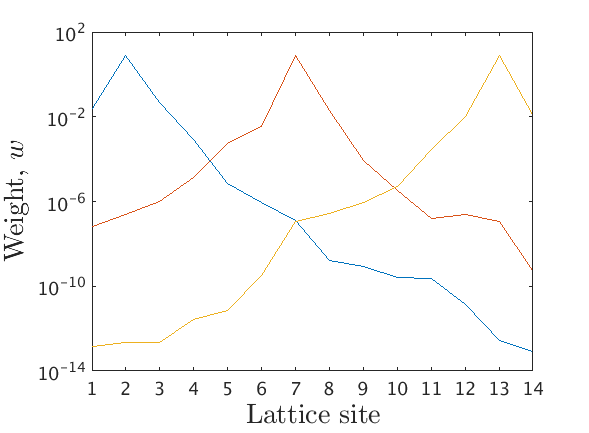}
	\caption[width=\linewidth]{Sample weights of three $\tau^x_i$ operators for a system of size $L=14$ and disorder $W=15$. The weight of the operator decays exponentially as a function of distance from the primary site.}\label{fig:txweights}
\end{figure}

In Ref. [\onlinecite{Kulshreshtha}], the authors of this paper presented a method to construct exact quasi-local integrals of motion. In this work, we present a new method to construct optimally local $l$-bit operators, called operator localization optimization (OLO). The method presented here improves on the previous method by reducing the runtime for systems of $L=14$ by approximately 40 times while improving operator localization length; on the computing cluster used by the authors (a 10 TFLOP Beowulf cluster of 113 multicore machines), this amounted to an absolute reduction in time from approximately ten hours to fifteen minutes.

To begin, we note that any operator $O$ can be written
\begin{equation}
O = \sum_{\gamma \in {0,x,y,z}} \sigma^{\gamma}_j \otimes A^{\gamma}_{\bar{j}}
\end{equation}
where $\sigma^0_i$ is the identity operator and $A^{\gamma}_{\bar{j}}$ is an operator that acts as the identity on site $j$ but is nontrivial elsewhere. The action of the operator on a site $j$ can be quantified as the weight function
\begin{equation}
\begin{split}
w_j (O) &= \frac{16}{\mathcal{N}} \Tr[(A^{x}_{\bar{j}})^2 + (A^{y}_{\bar{j}})^2 + (A^{z}_{\bar{j}})^2] \\
&= \frac{1}{\mathcal{N}} \sum_{\gamma \in {x,y,z}} \Tr[(O - \sigma^\gamma_j O \sigma^\gamma_j)^2]. 
\end{split}
\end{equation}
This non-negative function measures how much the operator $O$ affects site $j$, and it is zero if $O$ acts as the identity on site $j$. The normalization factor $\mathcal{N}$ is simply the size of the Hilbert space; in this case $\mathcal{N} = 2^L$.

Consider an $l$-bit labelling on the set of eigenstates $\{\alpha\}$. Each eigenstate is uniquely labeled by the chain of eigenvalues from the set $\{\tau^z_j \},$ for example $\{ + + - + - \}$ for a system with $5$ $l$-bit operators. The function $p(\alpha,i)$ takes an eigenstate $\alpha$ to the eigenstate with a label flipped on $i$ and identical everywhere else. For example, $p(\{+ + + +\},1) = \{- + + +\}$. We also define a measurement operator $s(\alpha,i)$ that takes the sign of the $l$-bit label of an eigenstate $\alpha$ on site $i$. For example, $s(\{- + + +\},1) = -1$. The $l$-bit measurement and flip operators can then be written
\begin{align}
\tau^z_i &= \sum_{\alpha} s(\alpha,i) \ket{\alpha} \bra{\alpha} \\
\tau^x_i &= \sum_{\alpha} \ket{\alpha} \bra{p(\alpha,i)}.
\end{align}
The functions $s$ and $p$ correspond to an $l$-bit algebra on eigenstates and equivalently a choice on the ordering of the columns of $U$.

For these operators, the weight function can be written as follows:
\begin{equation} \label{eq:tzweight}
\begin{split}
\mathcal{N} w_j (\tau^z_i) &= 3 \cdot 2^{L+1} \\ & \quad -2\sum_{\gamma \in {x,y,z}} \sum_{\alpha, \beta} s(\alpha,i) s(\beta,i) \bra{\alpha} \sigma^\gamma_j \ket{\beta} \bra{\beta} \sigma^\gamma_j \ket{\alpha} \\
&= 3 \cdot 2^{L+1}-2\sum_{\gamma} \Tr \left[ M^{\gamma,j}  \sigma^z_i M^{\gamma,j} \sigma^z_i \right]
\end{split}
\end{equation}
\begin{equation} \label{eq:txweight}
\begin{split}
\mathcal{N} w_j (\tau^x_i) & = 3 \cdot 2^{L+1} \\ & \quad - 2 \sum_{\gamma \in {x,y,z}} \sum_{\alpha,\beta} \bra{\alpha} \sigma^\gamma_j \ket{\beta} \bra{p(\beta,i)} \sigma^\gamma_j \ket{p(\alpha,i)} \\
&= 3 \cdot 2^{L+1} - 2\sum_{\gamma} \Tr \left[ M^{\gamma,j} \sigma^x_i M^{\gamma,j} \sigma^x_i \right]
\end{split}
\end{equation}
where $M^{\gamma,j}=U^\dagger \sigma^\gamma_j U$.

For a quasilocal operator on site $i$, $w_j \propto e^{|i-j|/\xi}$, where $\xi$ is the localization length of the operator. Therefore, our goal is to maximize the sums shown at the end of equations (\ref{eq:tzweight}) and (\ref{eq:txweight}) away from site $i$. Each term in the sum has a maximum $2^L$. Therefore, maximizing these will bring the total weight as close to $0$ as possible. As we expect the weight functions for $\tau^z_i$ and $\tau^x_i$ operators to mirror one another, we choose to focus just on the sum in equation (\ref{eq:txweight}), as we find that the weight decay of the $\tau^x_j$ operator is more sensitive to the pairing structure chosen.

The first insight in solving this optimization problem comes from noting that in a system without spin-spin interaction (in which the eigenstates are simply product states), the ideal ordering of $U$ is one such that $M^{\gamma,j} = \sigma^\gamma_j$. If this is the case, then for all $j \neq i$, the $\sigma^x_i$ operators in equation (\ref{eq:txweight}) commute through $M^{\gamma,j}$, yielding
\begin{equation}
\begin{split}
 \Tr \left[ M^{\gamma,j}  \sigma^x_i M^{\gamma,j} \sigma^x_i \right] = \Tr \left[ \sigma^x_i \sigma^x_i M^{\gamma,j} M^{\gamma,j}  \right] = 2^L .
\end{split}
\end{equation} 
Inserted back into equation (\ref{eq:txweight}), this yields $w_j(\tau^x_i) = 0$ as desired. For $i=j$, $M^{x,i}$ commutes through, yielding $2^L$ for this part of the sum as well. For the other parts of the sum, $M^{\gamma,i}$ anticommutes with $\sigma^x_i$, yielding $-2^L$ for each part of these sums. Thus, we obtain $w_i(\tau^x_i) = 8 \cdot 2^L$, the maximum allowed weight.

After turning on the spin-spin interaction, the same principle applies, though the ideal ordering becomes harder to find. We can minimize the weight where $j \neq i$ and maximize it where $j=i$ by satisfying the following two principles that can be seen from equations (\ref{eq:tzweight}) and (\ref{eq:txweight}).

\begin{enumerate}

\item For (a) $j \neq i, \gamma=x,y,z$ or (b) $j = i,\gamma = x$: if $|\bra{\alpha} \sigma^{\gamma}_j \ket{\beta}|$ is close to unity, then $\bra{p(\beta,i)} \sigma^{\gamma}_j \ket{p(\alpha,i)} \approx \bra{\alpha} \sigma^{\gamma}_j \ket{\beta}$. In case (a), the first line of equation (\ref{eq:txweight}) shows that the sum can be brought close to $2^L$, yielding a small value for $w_j(\tau^x_i)$. Case (b), for which $j=i$ is covered below.

\item For $j=i,\gamma = y,z$: if $|\bra{\alpha} \sigma^{\gamma}_i \ket{\beta}|$ is close to unity then $\bra{p(\beta,i)} \sigma^{\gamma}_j \ket{p(\alpha,i)} \approx -\bra{\alpha} \sigma^{\gamma}_i \ket{\beta}$, or vice versa if the first quantity is negative. For $\gamma=x$ (case 1b), the sum in the first line of equation (\ref{eq:txweight}) is close to $2^L$. For $\gamma = y,z$, the sum is close to $-2^L$. When inserted into the first line of equation (\ref{eq:txweight}), the total is brought to 8 $\mathcal{N}$, the maximum allowed value.

\end{enumerate}

These two principles, if satisfied, thus ensure that the weight of an operator $\tau^x_i$ is maximized at site $i$ and minimized elsewhere. However, the correct ordering is necessary in order to satisfy both principles.

Notice that (1b) and (2) are clearly satisfied by stipulating that if $\bra{\beta} \sigma^x_i \ket{\alpha} \gg 0$, then $p(\alpha,i) = \beta$. We will take this as an ansatz and numerically verify that the first condition is also satisfied.

As the Hamiltonian conserves total spin in the $z$ direction, its eigenstates can be split into sectors according to total spin. There are $L+1$ sectors that we will label $U_1 , U_2, \ldots , U_{L+1}.$ Utilizing conservation of total spin serves two purposes. First, splitting the Hamiltonian into spin sectors allows us to diagonalize smaller matrices and thereby work with larger systems. Second, the sectors give us some information on the $l$-bit algebra; the $\tau^i_x$ operator changes the total spin of an eigenstate by one unit. Therefore, $\ket{p(\alpha,i)}$ exists in an adjacent spin sector to an eigenstate $\ket{\alpha}$ for all $\alpha$ and $i$. This fact narrows down the search for $p(\alpha,i)$ from the full spectrum of eigenstates to just one or two spin sectors.

We now describe the pairing process inductively, starting from an eigenstate spin sector $U_i$ that we assume already has the correct $l$-bit labelling, meaning that its columns have been correctly ordered. Our goal is now to correctly order $U_{i+1}$. We start by taking the set of matrices $\{ O_j \} = \{ U_i^\dagger \sigma^+_j U_{i+1} \}$ over all $j$. 

The proper interpretation of each $O_j$ operator is as follows. The matrix $U_i^\dagger$ is one whose rows are eigenstates with a correct $l$-bit label. We then apply the operator $\sigma^+_j$ on the right, flipping the $j^\text{th}$ physical spin of each state from down to up or eliminating terms that are already up. Notice that $\sigma^+_j$ is a block off-diagonal matrix in the basis of total spin, with only upper triangular nonzero terms. This is the segment that flips $U_i$ into $U_{i+1}$ rather than $U_{i-1}$. This multiplication yields a matrix whose rows are the eigenstates from the $i^\text{th}$ spin sector with the $j^\text{th}$ physical spin flipped, though solely the part that lives in the $(i+1)^\text{th}$ spin sector. We then label the rows, flipping the $j^\text{th}$ $l$-bit. Taking the product of this matrix with $U_{i+1}$, whose columns are randomly arranged eigenstates. $O_j$ is thus an overlap matrix between $U_{i+1}$ and $U_{i}$ with a physical spin flip on site $j$.  

We now have a set of overlap matrices whose rows contain the set of labellings for the $(i+1)^\text{th}$ sector. Because there are $L$ such matrices, labellings may be represented multiples times, representing multiple approximations of the same eigenstate in the $(i+1)^\text{th}$ sector using different physical spin flips from the $i^\text{th}$ sector. If this is the case, we simply average the absolute value of the overlaps with identical row labellings. 

Our goal is now to match each column to a row labelling. We do this through a greedy algorithm, choosing each subsequent pairing by minimizing the second-worst available overlap of remaining eigenstates. This greedy algorithm generally outperforms simply repeatedly choosing the maximum available overlap.

The final important factor in assigning eigenstates is determining the appropriate phase. The phase, plus or minus one, of an eigenstate has a strong effect on the localization length of the $\tau^x_j$ matrix. As can be seen in equations (\ref{eq:tzweight}) and (\ref{eq:txweight}), the weight of an operator $\tau^x_i$ or $\tau^z_i$ outside of site $i$ is minimized if the terms $\bra{\alpha} \sigma^{\gamma}_i \ket{\beta}$ have the same sign, so we multiply the eigenstate in $U_{i+1}$ by the appropriate phase (plus or minus one) to make all of the overlap terms positive. 

We start this inductive process from $U_{0}$, which consists only of the all down physical spin eigenstate which we automatically assign the label $\{ - - - \ldots \}$, and iterate through spin sectors until reaching $U_{L+1}$, which consists only of the all up physical spin eigenstate.

The OLO algorithm gives us a good labelling of eigenstates to create localized $\tau^z_i$ and $\tau^x_i$ operators. The weight profiles of sample $\tau^x_i$ operators are given in Figure \ref{fig:txweights}.

\subsection{Extension of operators to large systems and measurement of local observables}
The OLO algorithm allows us to construct exact $l$-bits on one dimensional systems up to size $L=14$ using 16GB RAM. Attempting to access exact eigenstates and measure observables exactly on larger systems becomes computationally infeasible.

However, the quasilocal nature of the $\tau^j_x$ operators in FMBL systems presents a natural method to approximately move between eigenstates of large MBL systems. The action of an operator $\tau^j_x$ becomes exponentially trivial far away from the site $j$. Therefore, if we construct $\tau^j_x$ on an appropriate window, we can extend the operator to a larger FMBL system simply by extending it using the identity matrix. If the window used is large enough, the approximate operator only misses an exponentially decaying action outside the window. From a known eigenstate (such as the all up physical spin state), we can then use combinations of $\tau^j_x$ operators to approximately access other eigenstates. The $2^L$ different combinations of products of $\tau^j_x$ operators yield the $2^L$ distinct eigenstates. We call this eigenstate formulation the $\tau_x$ network representation.

Because the Hilbert space involved in this calculation grows exponentially with system size, it is not possible to store the full eigenstate, but it is possible to measure local observables on the eigenstates by selectively extending and tracing on the system such that the operating space is never too large to conduct calculations.

Starting from a large Hamiltonian of size $L$, we break the system down into 'manageable' sub-systems of size $l$. Where the original Hamiltonian is 
\begin{equation}
H = \sum_{i=1}^{L-1} \textbf{S}_i \cdot \textbf{S}_{i+1} + \sum_{i=1}^L h_i S^z_i,
\end{equation}
we create $L-l+1$ subsystems with Hamiltonian
\begin{equation}
H^{sub}_j = \sum_{i=j}^{j+l-2} \textbf{S}_i \cdot \textbf{S}_{i+1} + \sum_{i=j}^{j+l-1} h_i S^z_i.
\end{equation}
We use the algorithm described in the previous subsection to build a set of quasilocal $\tau_x^j$ operators on each site of each subsystem. 

\begin{figure}
	\includegraphics[width=\linewidth]{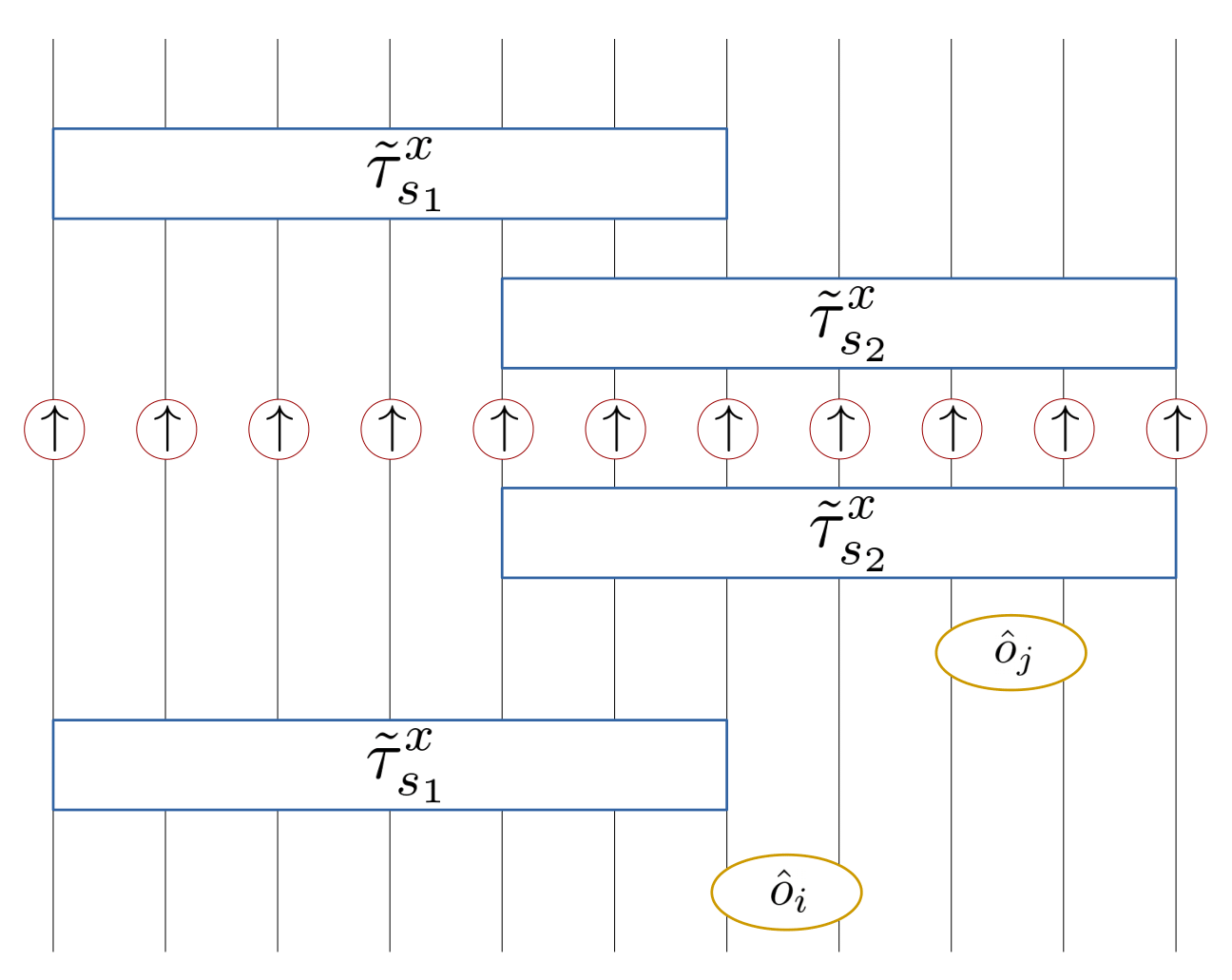}
	\caption[width=\linewidth]{A diagram portraying the structure of our $\tau^x$ network representation of eigenstates. Starting from the reduced density matrix $\rho_\uparrow$, represented in the red circles here, we carry out the multiplication from the inside outward. In order to implement our algorithm, we rearrange commuting operators so that this multiplication can be carried out from right to left, as indicated in equations (\ref{eq:observe1}) and (\ref{eq:observe}). In this case, $\left\langle \hat{o}_i \hat{o}_j \right\rangle = Tr(\tilde{\tau}^x_{s_1} \tilde{\tau}^x_{s_2} \rho_\uparrow \tilde{\tau}^x_{s_2} \tilde{\tau}^x_{s_1} \hat{o}_i \hat{o}_j).$ Note that $\hat{o}_j$ commutes through $\tilde{\tau}^x_{s_1}$ and can therefore be moved forward in the multiplication order, represented in this diagram by being moved inward. Operators $\hat{o}_i$ and $\tilde{\tau}^x_{s_1}$ however do not commute. Therefore, before including $\hat{o}_i$, we must extend the working window to include $\tilde{\tau}^x_{s_1}$, the largest working window we will need.}\label{fig:network}
\end{figure}

\begin{figure}
	
	\begin{minipage}{\linewidth}
		\includegraphics[width=\linewidth]{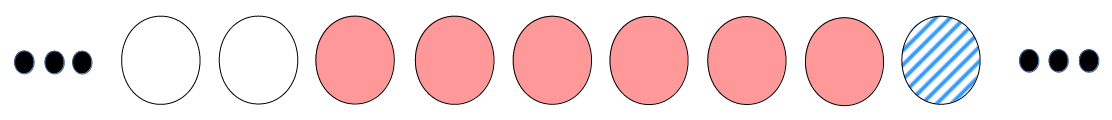}
		\subcaption{\label{fig:inchworm1}}
	\end{minipage}
	\begin{minipage}{\linewidth}
		\includegraphics[width=\linewidth]{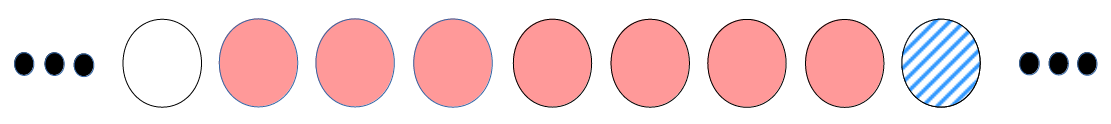}
		\subcaption{\label{fig:inchworm2}}
	\end{minipage}
	\begin{minipage}{\linewidth}
		\includegraphics[width=\linewidth]{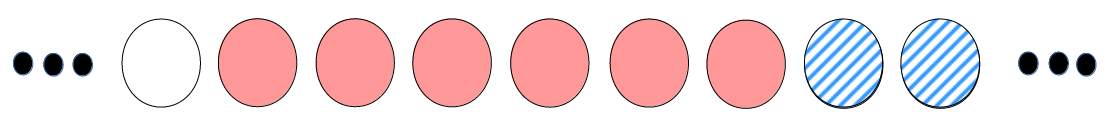}
		\subcaption{\label{fig:inchworm3}}
	\end{minipage}
	\caption[width=\linewidth]{A depiction of the algorithm described in the text. The white sites are those that have yet to be reached, the solid red sites are in the current working window, and the striped blue sites have been traced out. In (\subref{fig:inchworm1}), we start with a working window of size $l$. In (\subref{fig:inchworm2}), the working window extends one site to the left by taking a tensor product with the single-site spin up density matrix. The window is now of size $l+1$. If a $\tau^x$ operator or an observable exists in the window, it is introduced in this step. In (\subref{fig:inchworm3}), the working space contracts by taking the partial trace over the site on the right. The system is traversed in this manner. }\label{fig:inchworm}
\end{figure} 

For each site of the large system, we can choose a subsystem containing the site and a $\tilde{\tau}_x^j$ operator of size $l$ centered on the site. Generally, there will be $l$ subsystems containing any given site of the total system and ideally we select a $\tilde{\tau}_x^j$ operator from the center of a subsystem so as to cut off as little of the operator's action as possible. 

\begin{figure*}
	\includegraphics[width=\linewidth]{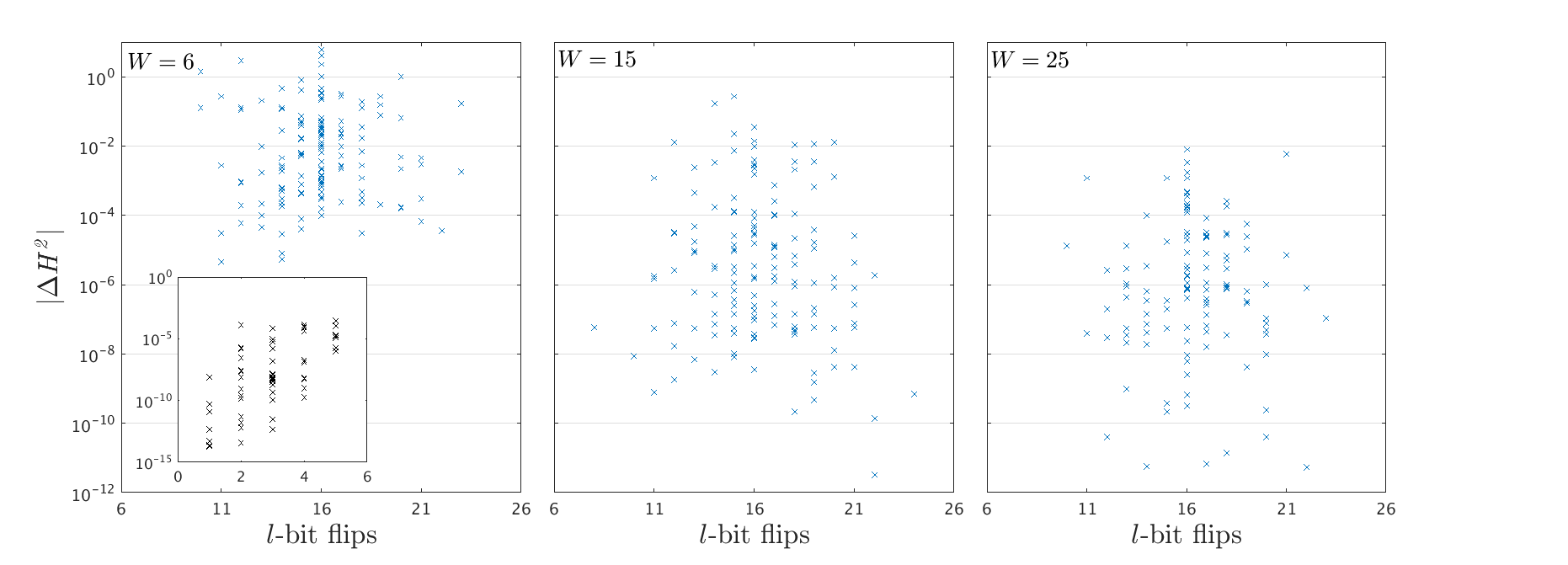}
	\caption[width=\linewidth]{Energy fluctuations plotted as a function of number of $l$-bit flips from the all up physical spin eigenstate for systems of size $L=32$. Three different disorder strengths are shown, with three realizations for each disorder strength and approximately fifty eigenstates per realization. If the number of bit flips is greater than $16$, the algorithm is started from the all down physical eigenstate instead. As expected, fluctuation increases with an increasing number of bit flips, as the approximate operators introduce error. Also as expected, fluctuation decreases with increasing disorder.}\label{fig:varvflips}
\end{figure*}

In practice, we select a $\tilde{\tau}_x^j$ operator by calculating the energy fluctuation of the eigenstate produced by applying each of the candidate operators from the $l$ subsystems containing the site $j$ to the fully polarized eigenstate. We select the candidate operator that produces the lowest fluctuation. Energy fluctuation is calculated using the process described below.

In our Hamiltonian, we know two eigenstates independent of the disorder realization: the all up and all down physical spin states ($\ket{\uparrow \uparrow \uparrow \ldots}$ and $\ket{\downarrow \downarrow \downarrow \ldots}$) We label these $\ket{+ + + \ldots}$ and $\ket{- - - \ldots}$ in the $l$-bit basis respectively. From these eigenstates, the set of $\tau^j_x$ operators that we have selected can be used to flip $l$-bits to attain any configuration, allowing us to target any eigenstate through its $l$-bit label.

We now proceed with a verbal and pictorial description of the process by which we calculate the expectation value of a product of local operators on a large system while working in a Hilbert space of computationally manageable size. Our starting eigenstate here is the all up physical spin state whose density matrix we label $\rho_\uparrow$. We additionally generate an $l$-bit flip configuration $S = \{s_1, s_2, \ldots s_n \}$ and a product over local observables $\hat{O} = \hat{o}_1 \hat{o}_2 \ldots \hat{o}_m $. The quantity to calculate is:
\begin{equation} \label{eq:observe1}
\begin{split}
\left\langle \hat{O} \right\rangle &= \Tr (\rho O)\\
&= \Tr \left[ \left( \prod_{j \in S} \tau^x_j \right) \ket{\uparrow \cdots} \bra{\uparrow \cdots} \left( \prod_{j \in S} \tau^x_j \right)^\dagger \hat{O} \right].
\end{split}
\end{equation}

We are careful to determine a canonical ordering of the product over $\tilde{\tau}^x_j$ operators; while the exact $\tau^x$ operators on the full system commute, ours may not commute exactly as they are not exact. We choose to order the operators in ascending order of $j$. Writing the products explicitly, our equation becomes
\begin{equation} \label{eq:observe}
\left\langle \hat{o}_1 \ldots \hat{o}_m \right\rangle = \Tr (\tilde{\tau}^x_{s_1} \ldots \tilde{\tau}^x_{s_n} \rho_\uparrow \tilde{\tau}^x_{s_n} \ldots \tilde{\tau}^x_{s_1} \hat{o}_1 \ldots \hat{o}_m).
\end{equation}
A visual representation is shown in Figure \ref{fig:network}.

Each of the $\tilde{\tau}^x_i$ operators has non-trivial support on a window of size $l$ and is trivially the identity outside of this window. Our goal is to never work with a reduced density matrix larger than the window. We initialize the process using the reduced density matrix of the all up physical spin eigenstate on the rightmost window. We then expand the system leftward, extending the window by taking the tensor product of an up spin with the current reduced density matrix. To keep the working space manageable, we subsequently contract the system from the right by taking the partial trace over the last site. For the process portrayed in Figure 
\ref{fig:inchworm} wherein the working window extends leftward and contracts rightward, we call this method the inchworm algorithm.

The order in which we introduce the operators is important, as it pertains to the order of multiplication. We always introduce $\tilde{\tau}^x_i$ operators when we reach the operators' right edge, indicating that we introduce the $l$-bit flip operators in descending order. When we reach the right edge of an observable, we must first introduce any $\tilde{\tau}^x_i$ operator that intersects with the observable in order to maintain the order of the operation in Equation (\ref{eq:observe}). 

For example, when we encounter an observable $o_j$ that has an intersecting range with a bit flip operator $\tilde{\tau}^x_i$, we first include the $l$-bit flip operator even if its support does not extend as far to the right as the support of the observable. We update the density matrix by
\begin{equation}
\rho_n = \tilde{\tau}^x_i (\rho_{\uparrow,s} \otimes \rho_0) \tilde{\tau}^x_i \hat{o}_j,
\end{equation}
 where $\rho_{\uparrow,s}$ is here the density matrix of all up physical spins of the length required to extend the $\rho_0$ to the range of $\tilde{\tau}^x_i$ and $\hat{o}_j$. Therefore, if $l$ is the lattice size of the $\tau^x_i$ operator and $l_o$ is the lattice size of the $\hat{o}_j$ operator, the largest window we ever need to work with has lattice size $l + l_o - 1$.

Progressing until we reach the furthest left site on the system and tracing over the remaining sites, we obtain the product in Equation (\ref{eq:observe}). A visual representation of this ordering is shown and explained in Figure \ref{fig:network}.

\section{Results}

\label{sec:results}

\subsection{Energy Fluctuation}

A natural first test to verify the quality of the algorithm described above is to calculate the energy fluctuation of approximate eigenstates produced using the method. Calculating the energy fluctuation has the benefit of indicating the quality of the eigenstates constructed by the approximate $l$-bit flip operators, thereby proving the use of the method for general products of observables.

The energy fluctuation or variance, $\Delta H^2 = \langle H^2 \rangle -  \langle H \rangle ^2$, can be calculated by splitting the Hamiltonian into a sum of local operators acting over two sites. In this case,
\begin{equation}
\begin{split}
H &= \sum_{i=1}^{L-1} \hat{h}_i , \\
\hat{h}_i &= \textbf{S}_i \cdot \textbf{S}_{i+1} + h_i S^z_i
\end{split} 
\end{equation}
We can then calculate the fluctuation by taking the sum over a set of products of local observable operators.

Figure \ref{fig:varvflips} shows the variance of eigenstates produced by different combinations of $l$-bit flips for three disorder strengths for systems of size $L=32$. Each disorder strength contains three disorder realizations and approximately fifty eigenstates for each realization.

Flipping no $l$-bits whatsoever, we expect a variance of zero, as the all up and all down physical spin states are exact eigenstates. We expect the variance to increase with number of bit flips because the approximate $l$-bit flip operators introduce error into the constructed eigenstate. Eigenstates are selected at random; therefore because the eigenstates follow a binomial distribution in number of $l$-bit flips, the number of bit flips is clustered about $L/2$. If the number of bit flips is greater than $L/2$, we start from the all down physical spin eigenstate, meaning that we never need to flip more than $L/2$ $l$-bits.

\begin{figure}
	
	\begin{minipage}{\linewidth}
		\includegraphics[width=\linewidth]{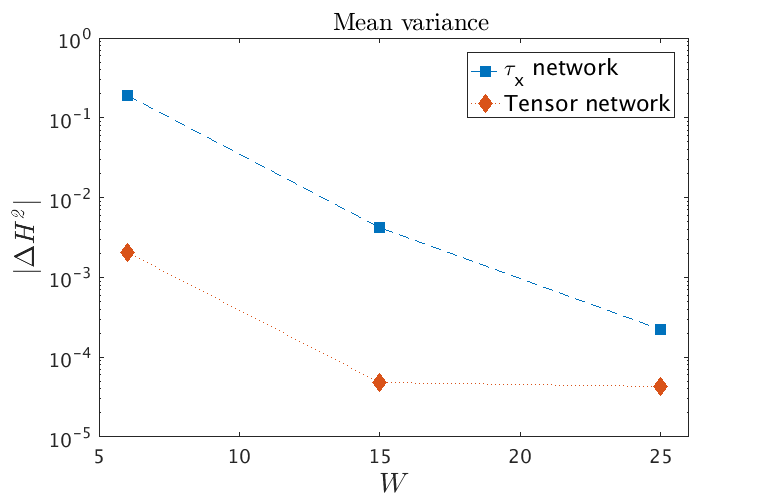}
	\end{minipage}
	\begin{minipage}{\linewidth}
		\includegraphics[width=\linewidth]{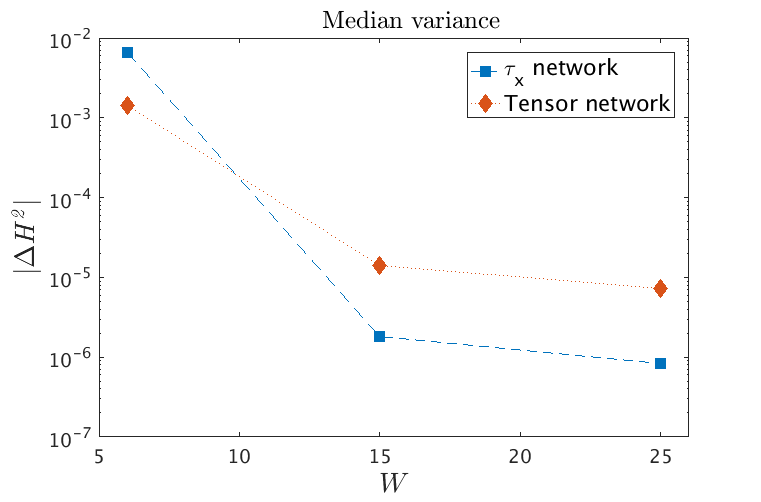}
	\end{minipage}
	\caption[width=\linewidth]{A comparison of the $\tau^x$ network representation presented in this paper with the tensor network representation for approximate eigenstates, both for systems of size $L=32$. \textit{(Top)} The mean of the $\tau^x$ network formulation is consistently worse, likely owing to the fact that the distribution of variances is uniform on a log scale and the mean is therefore dominated by upper outliers. \textit{(Bottom)} However, a comparison of the $50^{\text{th}}$ percentile variance of each method indicates that the median $\tau^x$ network eigenstate is more accurate than the median tensor network eigenstate for higher disorder strengths, $W \gtrapprox 10$. }\label{fig:both_comp}
\end{figure}

As a comparison to existing methods for approximating eigenstates on large localized systems, Figure \ref{fig:both_comp} shows the median and mean variance of eigenstates produced using the $\tau^x$ network described in this paper and the tensor network method described in Ref. [\onlinecite{Wahl}], where the constituting unitaries are of size $l=8$. The data are shown over three disorder strengths for systems of size $L=32$. For each disorder strength, three realizations are generated, with one hundred eigenstates per realization calculated using the $\tau^x$ network representation and one thousand for the tensor network method.

The mean for the $\tau^x$ network representation is consistently worse than that for the tensor network method owing to the fact that the $\tau^x$ network method's mean is dominated by outlying eigenstates of high variance (see Fig. 
\ref{fig:varvflips}). However, the median variance for the $\tau^x$ network method becomes lower than that for the tensor network method with increasing disorder strength. This indicates that deep in the MBL phase, the typical approximation yielded by the $\tau^x$ network representation becomes better than that yielded by the tensor network method. As disorder strength increases, the non-trivial portions of the exact $\tau^x_j$ operators cut out of the subsystem window become smaller as the operators become more local. As a result, our approximate $\tilde{\tau}^x_j$ operators resemble the exact $\tau^x_j$ operators to a higher degree with increasing disorder strength, yielding more accurate eigenstates.

\subsection{Correlations}

Our technique also allows us to probe long-range correlations of observables measured on eigenstates of MBL systems. Though the approximation of the $l$-bit flip operator cuts off operator weight outside of some window, approximate eigenstates composed of overlapping strings of $\tilde{\tau}^x_j$ operators can display correlation outside of this operator window length. Generally, two observables that can be continuously connected by windows of $\tilde{\tau}^x_j$ operators will display a non-trivial correlation. For systems of a moderate size such as the $L=32$ size systems with which we work, this condition is fulfilled for most eigenstates.

\begin{figure}
 	\includegraphics[width=\linewidth]{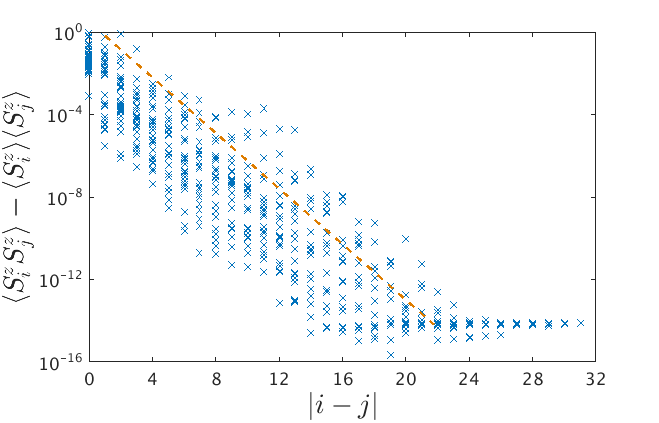}
 	\caption[width=\linewidth]{A plot of all correlations $\langle S^z_i S^z_j \rangle - \langle S^z_i \rangle \langle S^z_j \rangle$ for an approximate eigenstate of a system of size $L=32$ and disorder strength $W=6$. An exponentially decaying line approximately bounding the correlations is provided as a guide to the eye. The correlators decay with distance until they reach machine precision. Note further that this decay can still be observed beyond the subsystem window of size $l=14$.}\label{fig:correlation_ex} 
 \end{figure}

Because of the entanglement behavior of MBL eigenstates, local observables are thought to show an exponential decay in correlation with distance \cite{Imbrie2016a}. Though we could feasibly measure correlations between any local observable through our method, in this case we choose to focus on the spin-spin correlation function:
\begin{equation}
 \langle S^z_i S^z_j \rangle - \langle S^z_i \rangle \langle S^z_j \rangle.
\end{equation}
We expect
\begin{equation}
\text{max}(\langle S^z_i S^z_j \rangle - \langle S^z_i \rangle \langle S^z_j \rangle) \propto e^{|i-j|/\xi},
\end{equation} 
where $\xi$ is a localization length, indicating that correlations of an eigenstate as a function of distance are bounded by an exponentially decaying envelope. An example of this behavior for a low disorder eigenstate is shown in Figure \ref{fig:correlation_ex}. 
 
The $\tau^x$ network description of eigenstates is not immediately expected to be able to exhibit accurate correlation beyond the length of the subsystem. Overlapping $\tilde{\tau}^x$ operators can carry non-trivial action beyond the length of a subsystem, though it is not evident that this action is similar to that carried by products of exact $\tau^x$ operators. However, Figure \ref{fig:correlation_ex} shows that correlations continue to decay smoothly even outside of the subsystem window, which may indicate that the $\tau^x$ network eigenstate correlations are more accurate than expected.

\begin{figure}
	\includegraphics[width=\linewidth]{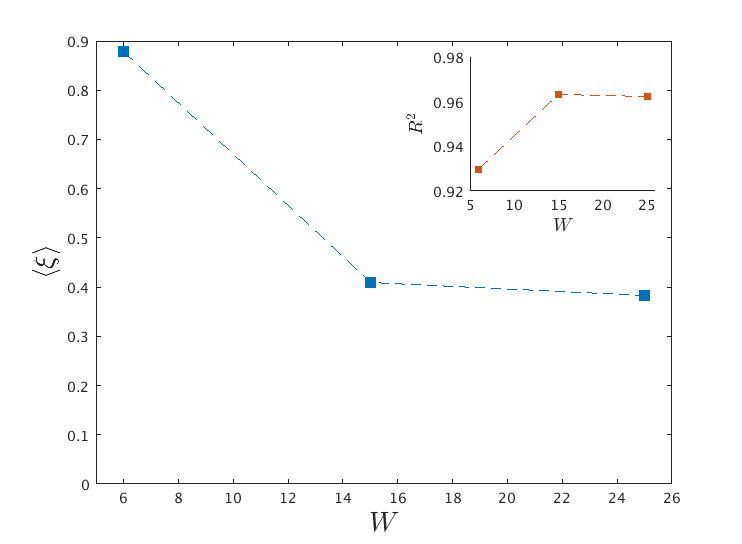}
	\caption[width=\linewidth]{The average value of $\xi$ as a function of disorder strength for one hundred eigenstates per realization and three realizations per disorder strength $W$. The value of $\xi$ is defined through the envelope bounding the correlation functions of an eigenstate: $(\langle S^z_i S^z_j \rangle - \langle S^z_i \rangle \langle S^z_j \rangle) \propto e^{|i-j|/\xi}$. Note that the localization length decreases with increasing disorder. In the inset, the quality of an exponential decay fit on the envelope is shown. The fit is of high quality for all $W$ considered in our simulations.} \label{fig:correlation_all}
\end{figure}

For each eigenstate calculated, approximately one hundred per disorder realization and three realizations per disorder strength, we calculated $\xi,$ the strength of the decay of the envelope of spin-spin correlation functions. The behavior of the mean $\xi$ as a function of disorder strength is shown in Figure \ref{fig:correlation_all}. Further, to determine the degree to which the correlator bound exhibits exponential decay, the average $R^2$ value for the exponentially decaying fit is shown in the inset.

As each of the disorder strengths tested are within the MBL phase, we do not observe a breakdown in the exponential decay of the correlator as a function of distance, even for the lowest disorder strength, $W=6$. However, we do observe a gradual increase of the exponential decay length with decreasing disorder strength as expected.

\section{Discussion and conclusions}
\label{sec:concl}

We presented in this work the $\tau^x$ network representation of approximate eigenstates and the inchworm method to measure observables on those eigenstates for large MBL systems. Benchmarked against the tensor network method, we found that the algorithm does not construct eigenstates as accurately as the tensor network method close to the MBL crossover. However, the median eigenstate constructed by the algorithm outperforms that produced by the tensor network method deep in the MBL phase.

In subsection \ref{subsec:current}, we outlined two current classes of methods to construct eigenstates on large MBL systems. Here, we briefly describe the advantages and disadvantages carried by the $\tau^x$ network representation compared to the others.

Like SIMPS, DMRG-X, and En-DMRG, the $\tau^x$ network representation allows one to construct highly excited eigenstates of MBL systems. Compared to the DMRG-X algorithm, the $\tau^x$ network formulation does not produce eigenstates as accurately. For example, the DMRG-X algorithm in Ref. [\onlinecite{Khemani2016}] produces eigenstates with mean error at machine precision for systems of size up to $L=40$ with disorder as low as $W=8$. However, one of the benefits carried by the $\tau^x$ network representation is that it does not rely on eigenstate overlap with a physical spin product state, allowing one to theoretically target any eigenstate by its $l$-bit label. By contrast, eigenstates with low overlap to physical spin product states may not be captured by DMRG-X.

The tensor network algorithm of Ref. [\onlinecite{Wahl}] also allows one to theoretically target any eigenstate by its $l$-bit label, making it most similar to the $\tau^x$ network representation. In terms of accuracy of eigenstate, the two algorithms are similar. The median tensor network eigenstate is an order of magnitude more accurate than that of the $\tau^x$ network formulation at $W=6$. However, at $W \gtrapprox 10$, the median $\tau^x$ network eigenstate becomes an order of magnitude more accurate than the tensor network algorithm.

A primary divergence between the tensor network and $\tau^x$ network algorithms comes from computational speed. At $L=32$, the tensor network algorithm can take a long period of time (on our 113-machine, 10 TFLOP Beowulf computing cluster up to a week) to generate the unitary matrix for a given realization, but thereafter, one can compute observables on eigenstates almost instantaneously. Meanwhile, the OLO algorithm presented in subsection \ref{subsec:olo} can generate an $l$-bit algebra for a system of size $L=32$ several times faster (on our computing cluster less than three hours), but the inchworm algorithm can take several hours to measure observables on a given eigenstate. Thus, the tensor network algorithm might be preferred when given a realization and a large set of eigenstates to sample. The $\tau^x$ network representation allows for a quick sampling over many realizations and, as opposed to DMRG-class methods, produces an unbiased sample of eigenstates.

There are several natural next steps in studying this algorithm. One would be to test it on other models with quasi-local operators that jump between eigenstates. For example, a recent work \cite{Wortis2017} explored quasi-local integrals of motion of a two-site, disordered Hubbard model by constructing $l$-bit-like algebras that could be used in the algorithm we present in this paper.  

Additionally, the algorithm structure presented in Figure \ref{fig:network} suggests that there is an analogy of this method to the tensor network algorithm. A future algorithm could use the $\tilde{\tau}^x$ operators from the OLO algorithm as a starting point, and then extend them by adding arbitrary unitary matrices on either side. These unitary matrices could then be optimized to minimize the commutation of the $\tilde{\tau}^x$ matrices or the energy fluctuation of the eigenstate produced by applying the $\tilde{\tau}^x$ matrices to the fully polarized eigenstate. 

\begin{acknowledgments} 
A.K.K. is supported by the Rhodes Trust. This project has received funding from the European Union’s Horizon 2020 research and innovation programme under the Marie Skłodowska-Curie grant agreement No. 749150. The contents of this article reflect only the authors' views and not the views of the European Commission. S.H.S. was supported by EPSRC grant EP/N01930X/1. Statement of compliance with EPSRC policy framework on research data: This publication is theoretical work that does not require supporting research data.
\end{acknowledgments}

\bibliography{bib2}

\end{document}